# Crystal structure of Nd$_{10.67}$Pt$_4$O$_{24}$, a new neodymium platinate

*Øystein Slagtern Fjellvåg[†], Helmer Fjellvåg[‡,*], Julie Hessevik[‡], Anja Olafsen Sjåstad[‡] and Gwladys Steciuk[§]*

[†] Department for Hydrogen Technology, Institute for Energy Technology, P.O. Box 40, Kjeller NO-2027, Norway;

[‡] Center for Materials Science and Nanotechnology, Department of Chemistry, University of Oslo, N-0315 Oslo, Norway,

[§] Institute of Physics of the CAS, Na Slovance 1999/2, 182221 Prague 8, Czech Republic

A new platinate was recently discovered when Nd$_2$O$_3$ was explored as a platinum capture material in the Ostwald process, formed by a direct reaction between PtO$_2$(g) and Nd$_2$O$_3$. The crystal structure of this new platinate and its composition, Nd$_{10.67}$Pt$_4$O$_{24}$, are here reported for the first time. The compound is synthesized either by a direct reaction using PtO$_2$(g) or by the citric acid chemical route. Based on 3-dimensional electron diffraction data and Rietveld refinement of high-resolution synchrotron and neutron powder diffraction data, we accurately describe its crystal structure in space group $I4_1/a$. The compound is structurally related to the $Ln_{11-x}$Sr$_x$Ir$_4$O$_{24}$ ($Ln$ = La, Pr, Nd, Sm) phases with a double-perovskite ($A_2BB'O_6$) like crystal structure and $A$-site cation-deficiency. Owing to the fixed oxidation state of Pt(IV), two of the four Nd-sites are partly occupied to provide charge neutrality, with the Nd4 site taking a split position. On heating, Nd$_{10.67}$Pt$_4$O$_{24}$ decomposes into Nd$_2$O$_3$ and Pt. A plateau in the thermogravimetric curves measured in 33 vol. % O$_2$ in N$_2$ indicates the presence of an intermediate Pt(II) phase at around 960 °C, probably isostructural with La$_4$PtO$_7$.

**Introduction**

The rare earths ($Ln$) cations are well known to form pyrochlore ($A_2B_2O_7$) type compounds together with tetravalent platinum, $Ln_2$Pt$_2$O$_7$ [1–4]. Such pyrochlores have attracted significant attention during the last decades owing to interesting physics of geometrically frustrated spin systems [1,5]. The stability of pyrochlores with trivalent $A$ and tetravalent $B$ cations is governed by the relative size of the $A$ and $B$ atoms. The stability window of $1.36 \leq R_A/R_B \leq 1.71$ can be extended when turning to high pressure synthesis routes [1]. Nd$_2$Pt$_2$O$_7$ [3,6] concurs with the $A_2B_2O_7$ stability criterion [1], but is only reported as a high pressure compound obtained at P = 40 kbar and T = 1620 K. The La-analogue falls outside the stability range and is not known. On the other hand, several detailed studies have been carried out for the heavier $Ln$-platinates like Gd$_2$Pt$_2$O$_7$ [7].

Although many rare earth platinates are conveniently synthesized by high pressure methods. Several platinates are already easily obtained by traditional solid-state reactions [8], e.g., double perovskites ($A_2BB'O_6$) with rare earth cations, $Ln_2B$PtO$_6$ ($Ln$ = La, Pr, Nd, Sm, Eu, Gd; $B$ = Mg, Co, Ni, Zn) [9], or by growth in high temperature carbonate fluxes, e.g., La$_3$NaPtO$_7$, Nd$_3$NaPtO$_7$ and La$_4$PtO$_7$ [10]. The direct reaction between the solid binary components, $Ln_2$O$_3$ and PtO$_2$, is hampered by the low thermal stability of PtO$_2$, being limited to around 600°C [8]. However, at high temperatures PtO$_2$ is reactive in its molecular form, and ternary platinates can be formed within an appropriate synthesis environment [11]. Gas streams containing PtO$_2$ can be achieved by passing air or oxygen over a heated Pt-filament. In this way volatile platinum species are transported in vapor phase and can react with oxides in their solid state to form platinates, see example reaction (1) and (2) [11-12].

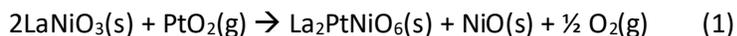

2LaNiO$_3$(s) + PtO$_2$(g) → La$_2$PtNiO$_6$(s) + NiO(s) + ½ O$_2$(g)     (1)

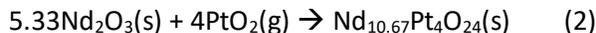

$$5.33 Nd_2O_3(s) + 4 PtO_2(g) \rightarrow Nd_{10.67}Pt_4O_{24}(s) \qquad (2)$$

An extensive study describes solid state synthesis of *Ln*-Pt-O compounds using mechanic mixtures of $Ln_2O_3$, $PtO_2$ and Pt metal wrapped in an Au foil and heat treated in sealed evacuated silica glass vials at 900 °C. In this way, $Ln_2PtO_4$ (*Ln* = La, Pr, Nd, Sm, Eu, Gd) were achieved, both notably with Pt(II) cations [8]. High temperature crystal growth in carbonate fluxes gave interestingly both Pt(II) compounds like $La_4PtO_7$ but also Pt(IV) compounds like $La_3NaPtO_7$ and $Nd_3NaPtO_7$ [10]. The different oxidation states for platinum have a distinct structural chemistry impact with respect to local coordination. The $d^8$ Pt(II) cations in $La_4PtO_7$ exhibit square planar coordination (Pt – O; 2 × 2.053Å; 2 × 2.045 Å) with $[PtO_3]^{4-}$ zigzag chains of corner-sharing square planar units along the *b*-axis. Similar structural features are found for a number of Pd(II) compounds, like $La_4PdO_7$ [10]. On the other, for Pt(IV) in $La_3NaPtO_7$ the $d^6$ Pt(IV) cations take regular octahedral coordination (Pt – O; 6 × 2.023Å) [10].

The basis for the current work is our observation that volatile $PtO_2$ species reacts at high temperatures with $Nd_2O_3$ under formation of a yet unknown compound. This was noticed during our efforts on capturing volatile Pt-species in high-temperature processes like the ammonia combustion process. During these experiments, $PtO_2$ was produced by passing air over a heated Pt-wire and transported in a quartz tube system before reacting with $Nd_2O_3$ pellets [11]. This new phase was thereafter synthesized by means of the citric acid wet-chemical route, which after optimization of the cationic composition, yielded a phase pure product of $Nd_{10.67}Pt_4O_{24}$. We currently report on the crystal structure of the non-stoichiometric $Nd_{10.67}Pt_4O_{24}$ compound, exhibiting a crystal structure similar to that of $Ln_9Sr_2Ir_4O_{24}$ [13]. Several isostructural compounds are known, including $A_{11}Re_4O_{24}$ (*A* = Ca, Sr) [14,15] and $Ba_{11}Os_4O_{24}$ [16]. The crystal structure was solved and refined on the basis of 3D ED (3-dimensional electron diffraction) [17,18], refined based on synchrotron powder diffraction data, and carefully evaluated with respect to neodymium non-stoichiometry and oxygen coordination by means of high-resolution powder neutron diffraction data. The oxidation state of platinum was evaluated based on thermogravimetric data for complete decomposition into $Nd_2O_3$ and Pt(s). This decomposition was furthermore carefully investigated by means of thermogravimetric analysis (TGA) and high temperature powder X-ray diffraction. The results are discussion in relation to the crystal structure of $Ln_9Sr_2Ir_4O_{24}$ and related compounds, as well as to other *Ln*-platinates.

**Experimental**

$Nd_{10.67}Pt_4O_{24}$ was synthesized using the wet-chemical citric acid complexation method. Starting materials were $Nd_2O_3$ (99.9 %, Sigma-Aldrich), Pt metal (99.9 %, K. A. Rasmussen), $C_6H_8O_7 \times H_2O$ (citric acid monohydrate, purity ≥ 99.5 %, Sigma-Aldrich), $O_2(g)$ (99.999 %, AGA), $HNO_3$ (65 wt. %, Merck KGaA) and HCl (37 %, Fischer Scientific). Prior to synthesis, $Nd_2O_3$ was annealed at 900 °C to remove any hydroxide and carbonate and then cooled in a desiccator. $Nd_2O_3$ was thereafter dissolved in 6M $HNO_3$ and Pt metal was dissolved in aqua regia ($HNO_3$:HCl = 1:3). The precursor solutions were mixed to synthesize a first batch with a nominal 3:1 stoichiometric ratio between Nd and Pt. Based on preliminary phase content analysis, a second batch was made with the exact composition of the target phase, $Nd_{10.67}Pt_4O_{24}$. Around 50 g citric acid was added to the solution per gram of $Nd_{10.67}Pt_4O_{24}$ product. The solution was heated during boiling with release of water and nitrous gases, followed by overnight heat treatment at 180 °C. Calcination was done at 400 °C in static air for 12 hours in a muffle furnace. The finely crushed powder was pressed to cylindrical pellets using a static pressure of 100 bar. Sintering was performed in a flow of 1 atm $O_2(g)$ at 800 °C. The process was repeated twice with duration times of 24 and 72 hours, respectively, with intermediate crushing and re-pelletizing.

For single crystal characterization by electron diffraction, the powder was dispersed in ethanol and deposited on a Cu grid coated by a holey amorphous carbon film. 3-dimensional electron diffraction (3D

ED) data were collected using a continuous mode (cRED) in a FEI Tecnai 02 TEM (acceleration voltage of 200 kV, $LaB_6$) equipped with a side-mounted hybrid single-electron detector ASI Cheetah M3, 512 × 512 pixels with high sensitivity and fast readout [17,18]. The data collection is automated by our in-house software *RATS* during which, series of non-oriented patterns are continuously collected by steps of 0.4° on the accessible tilt range. A dozen data sets were collected on different crystals to get an overview of the sample (Figure 1). The best data with the lowest *RC width* = 0.001 Å$^{-1}$ and *apparent mosaicity* = 0.0473 ° was selected for structure characterization (Figure 1 and Figure S1). 3D ED data reduction was performed with the program PETS2 [19–22] (Figure 2). The specific data processing for cRED data is extensively detailed in Klar *et al.* [22]. To model experimental intensities from continuous rotation data, Overlapping Virtual Frames (OVFs) are produced by summing consecutive experimental diffraction patterns into a set of virtual frames. Each OVF is characterized by its angular tilt range $\Delta\alpha_v$ covered by the virtual frame and the angular tilt step between two virtual frames (Table S1). The data reduction results in two *hkl*-types of files: one assuming the kinematical approximation later used in the structural solution (and the kinematical refinement) with *R(int)/wR(int)* = 0.3593/0.3607 and 92 % coverage for $\sin\vartheta/\lambda$ = 0.75 Å$^{-1}$ for the Laue class $4/m$, the other one dedicated to the dynamical refinement where each *OVF* is independently refined [23,24]. The structure was solved using Superflip [25,26] in Jana2020 [27] and refined using the dynamical theory with DYNGO in Jana2020. The data collection and refinement details are presented in the Supplementary Information.

Powder synchrotron X-ray diffraction data was measured at the Swiss-Norwegian Beamlines (SNBL; BM31) at the European Synchrotron Radiation Facility, Grenoble, using a wavelength of 0.25509 Å (0.3 mm glass capillary; transmission mode; $LaB_6$ calibration). The data was integrated using the Bubble software into 5000 data points with a step size of 0.004. The Rietveld analysis used the structure model as obtained by 3D-ED as starting point and was carried out using the JANA [27] suite of programs. Standard characterization in the home laboratory was done using a Bruker D8 Discover in reflection mode (CuK$\alpha_1$; Lynxeye detector) and in transmission (MoK$\alpha$, 2D Eiger detector) geometry.

Powder neutron diffraction time-of-flight (TOF) data were collected on a 2.5 g powder sample at the GEM instrument [28] at ISIS pulsed neutron and muon source, UK. The sample was mounted in a vanadium can and data was collected at room temperature. Instrumental parameters were obtained by refinement of a NIST Si standard. The diffraction data was analyzed using the TOPAS software [29]. In the final refinement we refined 2 lattice parameters, 26 atomic coordinates, 5 isotropic displacement parameters (individual for Pt, and one common for neodymium and one common for oxygen), and 2 occupancies (Nd3 and Nd4). In addition, instrumental parameters for each detector bank were refined.

Thermogravimetric analysis (TGA) was carried out by measuring data on a NETZSCH STA-449 F1 Jupiter unit using alumina crucibles. Experiments were carried out between room temperature and 1100 °C in a 33 vol. % $O_2$ in $N_2$ gas mixture or $N_2$ gas over the sample and by using a total flow-rate of 60 mL/min.

**Results and discussion**

**Crystal structure and structural chemistry**

The obtained grey-green powder of $Nd_{10.67}Pt_4O_{24}$ was phase pure when prepared with a citric acid chemical route. However, for the sample with nominal Nd:Pt ratio of 3:1, the Rietveld refinements revealed two small, unfitted peaks that corresponded neither to $Nd_2O_3$ nor $Nd_2Pt_2O_7$. As the crystal structure of the compound was unknown, we started our investigations by solving the crystal structure.

To identify the crystal structure of $Nd_{10.67}Pt_4O_{24}$, crystallites of the material were investigated by 3D ED. The indexing in PETS2 first offered a *F*-centered pseudo-cubic unit cell of about $a_{cubic}$ = 16.12(2) Å. A fine analysis of distortion parameters [20] together with the first integration statistics showed instead a tetragonal body centered unit cell with $a$ = 11.3474(1) Å and $c$ = 16.203(1) Å, Figure 1. The symmetry determination remained ambiguous from cRED as very strong dynamical effects hide the systematic extinctions, Figure 1c. Therefore, the synchrotron powder data were used to evaluate the space groups and refine the lattice parameters with better accuracy. Evaluation of the possible space groups shows that $I4_1/a$ provides the best fit, later confirmed by the symmetry test in SUPERFLIP.

The structure model was then obtained from cRED data in SUPERFLIP (Jana2020) and refined using the dynamical theory of electron diffraction. The dynamical refinement converged toward $R(obs)/wR(obs)$ = 0.0902/0.0962, $R(all)/wR(all)$ = 0.1045/0.1001 for 6991/9953 observed/all reflections and 158 refined parameters (Table 1). The atomic coordinates obtained are given in Table 2. The obtained composition $Nd_{10.733(16)}Pt_4O_{24}$, as refined from cRED data, shows a clear *A*-site cation deficient compound, which is structurally closely related to double perovskite iridates and consistent with the charge neutrality considering Nd(III) and Pt(IV). Both the site Nd3 and Nd4 sites show partial occupancy 0.988(6) and 0.379(5), respectively. Similar structures are reported for $Ln_9Sr_2Ir_4O_{24}$ (*Ln* = La, Pr, Nd, Sm) and related compounds [13–16]. The structural information from 3D-ED is summarized in Table 1.

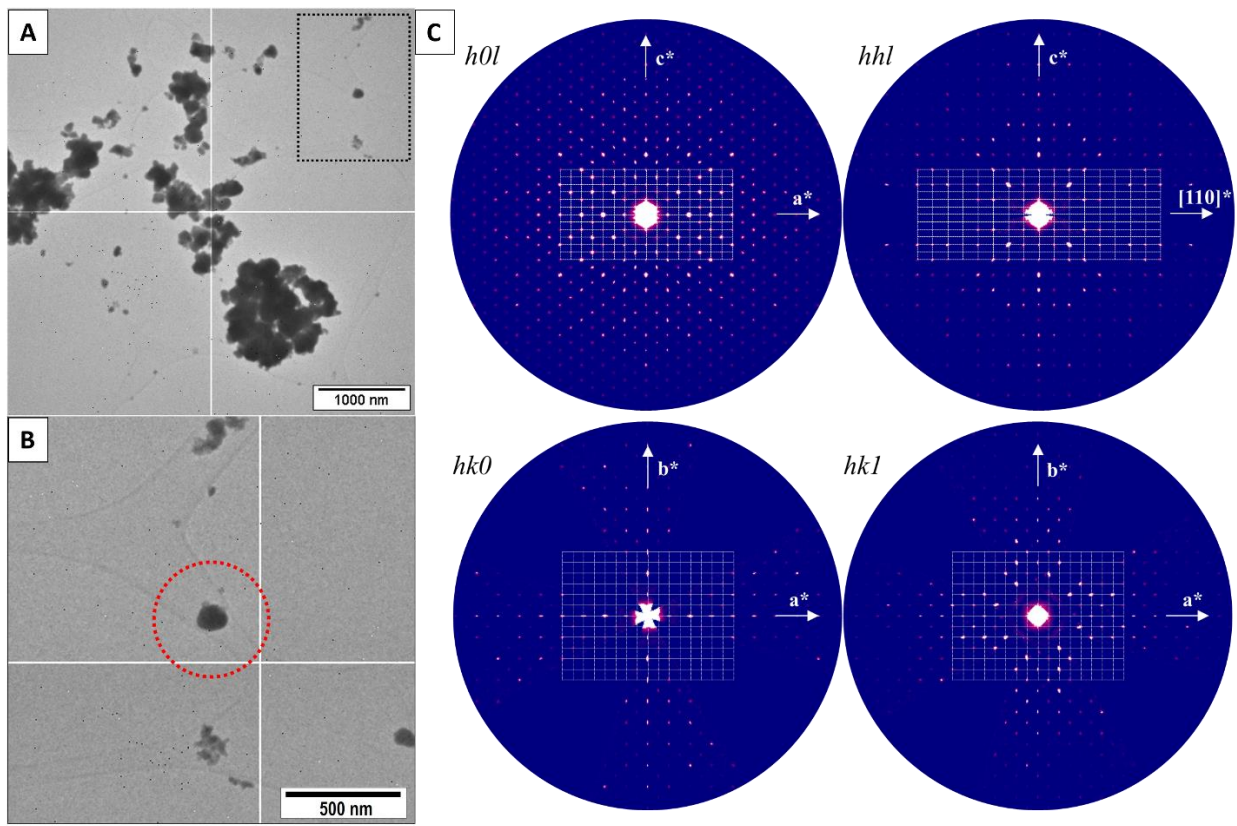

**Figure 1: A** – Overview of the sample's morphology under TEM. **B** – Crystal selected for 3D ED analysis with the nanobeam size of about 500 nm. **C** – Essential sections of the reciprocal space to define the symmetry of a tetragonal system.

**Table 1**. 3D ED data collection and structure refinement details. Details of Rietveld refinement of synchrotron powder data.

| | |
|---|---|
| Refined structural formula (dyn) | $Nd_{10.732}Pt_4O_{24}$ |
| *Crystal system* | tetragonal |
| *a* | 11.35203(11) Å |
| *c* | 16.21114(18) Å |
| *V* | 2089.11(4) Å$^3$ |
| *Z* | 4 |
| Density [g·cm$^{-3}$] | 8.6235 |
| Space group | *I*4$_1$/*a* |
| Temperature | ambient |
| TEM | FEI Tecnai 02 |
| Radiation (wavelength) | electrons (0.0251 Å) |
| Crystal dimensions (nm) | 140 x 100 |
| Δ*α*/total *α*-tilt (°) | 0.4/100 |
| *OVF:* Δ*α$_v$*/step between *OVF*(°) | 2.8/1.6 |
| Resolution range (θ) | 0.08–1.15 |
| Limiting Miller indices | *h*: –12→ 12, *k*: 0→18, *l*: 0 → 25 |
| No. of independent reflections (obs/all) – kinematic | 1906/2116 |
| *R*$_{int}$ (obs/all) – kinematic | 0.3593/0.3607 |
| Redundancy | 4.797 |
| Coverage for sinθ/λ = 0.8 Å$^{-1}$ | 92.12% |
| **Kinematical refinement of non-hydrogen atoms** | |
| *No. of reflections (obs/all) up to* sinθ/λ = 0.75 Å$^-$ | 1904/2114 |
| *Extinction correction -Becker* | *Giso = 0.474837* |
| *R, wR* (obs); *R, wR* (all); | 0.3137/0.3660; 0.3248/0.3690 |
| *N* refined param. | 37 |
| *formula* | Nd$_{10.733(16)}$Pt$_4$O$_{24}$ |
| *Occupancies:Nd3, Nd4* | 0.988(6), 0.379(5) |
| **Dynamical refinement** | |
| *No. of collected reflections (obs/all)* | 19225/26121 |
| *Selection criteria RSg(max)* | 0.6 |
| *No. of filtered outliers for* |*F$_{obs}$-F$_{calc}$*|>10σ(*F$_{obs}$*) | 27 |
| *Thickness model* | Wedge |
| *Effective thicknesses* | 1018(5) |
| *No. of reflections (obs/all)* | 6991/9953 |
| *GOF(obs)/ GOF(all)* | 0.0189/0.0166 |
| *R, wR* (obs) | 0.0902/0.0962 |
| *R, wR* (all) | 0.1045/0.1001 |
| *N* all param./*N* struct. parameters | 158/96 |
| **Rietveld refinement from synchrotron powder data** | |
| *No. of reflections (obs/all)* | 1244/1256 |
| *R, wR* (obs) | 0.0404/0.0677 |
| *R, wR* (all) | 0.0435/0.0677 |
| *No. of refined param. (structural ones)* | 35 |
| *Rp, wRp, GOF* | 0.0188, 0.0276, 0.1062 |
| *Refined formula* | Nd$_{10.796}$Pt$_4$O$_{24}$ |
| **Rietveld refinement from neutron powder data** | |
| *No. of reflections (obs/all)* | 14816/14981 |
| *No. of refined param. (structural ones)* | 32 |
| *Rexp, Rp, wRp, GOF* | 0.0204/0.0299/0.0210/1.0292 |
| *Refined formula* | Nd$_{10.696(1)}$Pt$_4$O$_{24}$ |

**Table 2:** Atomic coordinates and anisotropic displacement parameters for $Nd_{10.733}Pt_4O_{24}$ based on refinement of 3D ED data. Space group $I4_1/a$. Calculated standard deviations in parentheses.

| Atom | x | y | z | Ueq/Uiso | Occ | Wyckof |
|---|---|---|---|---|---|---|
| Nd1 | 0.72619(11) | 0.70294(10) | 0.50834(6) | 0.0121(3) | 1 | *16f* |
| Nd2 | 0.79388(11) | 0.47641(11) | 0.66284(6) | 0.0137(3) | 1 | *16f* |
| Nd3 | 1 | 0.5 | 0.49058(10) | 0.0213(5) | 0.988(6) | *8e* |
| Nd4 | 0.5 | 0.5 | 0.5342(2) | 0.0179(12) | 0.379(5) | *8e* |
| Pt1 | 0.75 | 0.5 | 0.375 | 0.0100(3) | 1 | *8c* |
| Pt2 | 1 | 0.75 | 0.625 | 0.0109(3) | 1 | *8d* |
| O1 | 0.9068(5) | 0.3242(5) | 0.7184(3) | 0.0128(14) | 1 | *16f* |
| O2 | 0.6503(6) | 0.6443(5) | 0.3686(3) | 0.0144(14) | 1 | *16f* |
| O3 | 0.8915(5) | 0.6161(5) | 0.5933(3) | 0.0132(12) | 1 | *16f* |
| O4 | 0.8788(5) | 0.6216(6) | 0.3986(3) | 0.0197(18) | 1 | *16f* |
| O5 | 0.5069(5) | 0.7078(6) | 0.5041(3) | 0.0159(15) | 1 | *16f* |
| O6 | 0.6654(5) | 0.8912(5) | 0.4519(3) | 0.0136(14) | 1 | *16f* |
| ADP harmonic parameters | | | | | | |
| Atom | U11 | U22 | U33 | U12 | U13 | U23 |
| Nd1 | 0.0163(6) | 0.0081(5) | 0.0119(4) | 0.0018(6) | -0.0016(4) | 0.0003(3) |
| Nd2 | 0.0120(5) | 0.0094(5) | 0.0197(5) | -0.0023(6) | -0.0035(4) | 0.0013(4) |
| Nd3 | 0.0261(11) | 0.0129(8) | 0.0248(9) | 0.0075(10) | 0 | 0 |
| Nd4 | 0.014(2) | 0.010(2) | 0.029(2) | -0.001(2) | 0 | 0 |
| Pt1 | 0.0082(6) | 0.0108(6) | 0.0111(5) | 0.0011(7) | -0.0010(4) | -0.0001(4) |
| Pt2 | 0.0119(6) | 0.0101(6) | 0.0108(5) | -0.0017(7) | 0.0000(4) | 0.0001(5) |
| O1 | 0.011(3) | 0.015(3) | 0.012(2) | 0.000(3) | -0.0045(18) | 0.0007(18) |
| O2 | 0.020(3) | 0.007(2) | 0.016(2) | 0.007(3) | 0.003(2) | 0.0036(18) |
| O3 | 0.006(3) | 0.016(3) | 0.018(2) | 0.002(3) | 0.0078(18) | 0.0025(19) |
| O4 | 0.017(3) | 0.025(4) | 0.017(2) | -0.012(3) | -0.004(2) | 0.007(2) |
| O5 | 0.010(2) | 0.028(3) | 0.010(2) | 0.002(3) | 0.0047(18) | -0.0007(19) |
| O6 | 0.014(3) | 0.008(2) | 0.019(2) | 0.001(3) | 0.004(2) | 0.0090(19) |

The single crystal structure obtained by 3D ED was evaluated at the powder scale by means of Rietveld refinements of the synchrotron powder diffraction data (Table 1). An excellent agreement between the model and the experimental data validated the structural model. The refinement details are given in Table 1 and the obtained atomic coordinates are given in the Supplementary Information Figure S2.

We note that the Nd4-atom is displaced out of the center of the (deformed) cuboctahedron and obtains thereby an improved bonding situation. The split position reflects hence a double-well potential for the best occupation of the Nd4-cation within the large void. To shed more light on the non-stoichiometry and the split position behavior suggested by 3D ED and X-ray diffraction, powder neutron diffraction data was collected. Specifically, the different contrast of neutrons compared to electrons and X-rays makes neutrons a better probe for oxygen positions, which provide means to unambiguously verify the oxygen environment along with the neodymium non-stoichiometry (Table 1).

First, the structural model from 3D ED and powder X-ray diffraction is in excellent agreement with the neutron data. By restricting the occupancy of Nd4 to unity while constraining all neodymium sites to have the same thermal displacement parameters, we observe some discrepancies in the fitted patterns and an $R_{wp}$ of 3.26 %. The discrepancies are reduced by turning to individual thermal displacement parameters, and in particular when refining the Nd4 occupation number. It is thus clear that the Nd4 site displays cation vacancies. By restricting all the neodymium sites to have the same thermal displacement parameters and allowing refinement of the Nd4 occupancy, we reach a final $R_{wp}$ of 2.09 %. The occupancy of Nd4 is 0.404(3). This corresponds to a final composition of $Nd_{10.808(6)}Pt_4O_{24}$, only slightly above the values obtained from 3D ED ($Nd_{10.733(16)}Pt_4O_{24}$) and synchrotron diffraction ($Nd_{10.796}Pt_4O_{24}$).

We further evaluated vacancies at the other Nd sites and find that the Nd3 site occupancy consistently converges to 0.947(4), with the $R_{wp}$ decreasing to 2.04 %. The final refinements thus have an occupancy of 0.947(4) and 0.401(3) on the Nd3 and Nd4 sites, respectively. This yields an overall composition of $Nd_{10.696(1)}Pt_4O_{24}$, which is in excellent agreement with values obtained from 3D ED and synchrotron diffraction. The composition is also close to what is expected from charge neutrality, namely $Nd_{10.667}Pt_4O_{24}$. The final structural model from neutron powder diffraction is given in Table 3, and the refinement of the third detector bank is showed in Figure 2. Refinements of the other detector banks and bond lengths are given in the Supplemental Information Figures S3-S7.

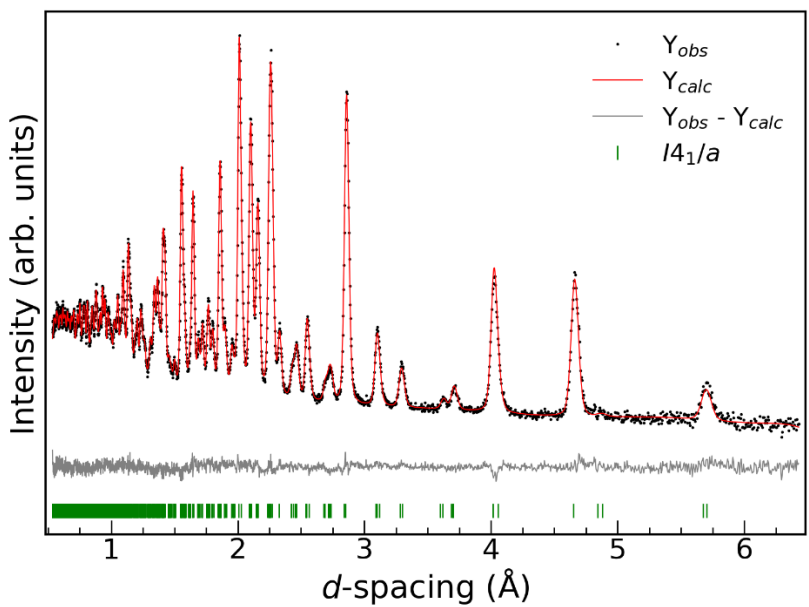

**Figure 2:** Rietveld refinement of neutron powder diffraction data from the third detector bank at GEM. The diagram shows the measured data (black dots), the calculated plot (red line), and the difference curve (gray). The Bragg reflection positions are given by green ticks.

**Table 3:** Atomic coordinates and isotropic displacement parameters for $Nd_{10.696(1)}Pt_4O_{24}$ based on refinement of neutron powder diffraction data. Space group $I4_1/a$ with lattice parameters $a$ = 11.3522(4) Å and $c$ = 16.2122(5) Å. Calculated standard deviations in parentheses.

| Atom | x | y | z | Biso (Å²) | Occ | Site |
|---|---|---|---|---|---|---|
| Nd1 | 0.72624(12) | 0.70380(11) | 0.50814(9) | 0.741(14) | 1 | *16f* |
| Nd2 | 0.79456(10) | 0.47697(12) | 0.66240(7) | 0.741(14) | 1 | *16f* |
| Nd3 | 0 | 0.5 | 0.48973(10) | 0.741(14) | 0.947(4) | *8e* |
| Nd4 | 0.5 | 0.5 | 0.5344(2) | 0.741(14) | 0.401(3) | *8e* |
| Pt1 | 0.75 | 0.5 | 0.375 | 0.293(19) | 1 | *8c* |
| Pt2 | 0 | 0.75 | 0.625 | 0.323(19) | 1 | *8d* |
| O1 | 0.90678(15) | 0.32564(17) | 0.71902(10) | 0.969(13) | 1 | *16f* |
| O2 | 0.65140(16) | 0.64422(17) | 0.36876(12) | 0.969(13) | 1 | *16f* |
| O3 | 0.89141(17) | 0.61478(17) | 0.59348(10) | 0.969(13) | 1 | *16f* |
| O4 | 0.88048(18) | 0.6205(2) | 0.39771(10) | 0.969(13) | 1 | *16f* |
| O5 | 0.5087(2) | 0.70718(15) | 0.50388(10) | 0.969(13) | 1 | *16f* |
| O6 | 0.66503(16) | 0.89043(18) | 0.45295(12) | 0.969(13) | 1 | *16f* |

The aspects of the correct positions for Nd4 and O5 sites were also evaluated based on the neutron powder diffraction refinements. If Nd4 were located at the *4a*-site (½, ½, ½) it would take a deformed cuboctahedral 12-fold coordination. However, the Nd–O bonds are in that case not favorable; four short 2.3549(18) Å, and eight long 3.0736(19) - 3.188(2) Å for Nd in the *4a* site, compared to the refined bond lengths given in Table 3 for the *8e* site. Hence, by shifting the Nd-cation out of the center, the coordination changes. Four of the twelve Nd-O bonds become significantly elongated, and a more favorable bonding situation can be achieved by shifting Nd4 in the vertical direction of the polyhedron, i.e. into the split-position. In the Rietveld refinements of the neutron data, when moving Nd4 from (½, ½, z) with z ≈ 0.53 to (½, ½, ½), the $R_{wp}$ increase to 2.4 %. We therefore reject that Nd4 is on the *4a* site and conclude that the *8e* site is correct.

**Table 4:** Obtained Nd4-O interatomic distances from neutron powder diffraction refinements.

| Atoms | Distance (Å) | Distance (Å) |
|---|---|---|
| Nd4-O2 | 2.846(3) × 2 | 3.584(4) × 2 |
| Nd4-O4 | 2.662(3) × 2 | 3.525(4) × 2 |
| Nd4-O5 | 2.4053(19) × 2 | 2.434(2) × 2 |

The O5 site is close to having coordinates (½, y, ½), however (½, y, ½) is not a special position, it is still the *16f* site. Moving O5 from (0.5087(2), 0.70718(15), 0.50388(10)) to (½, y, ½) gave a slight reduction in the quality of fit and gave $R_{wp}$ = 2.27 % and simultaneously increased the thermal displacement parameter of O5 from about 1 to 1.69 Å², i.e. 50 % higher than for the rest of the oxygen positions. We thus conclude that the coordinates of O5 must be freely refined, and that Nd4 takes a split position.

The crystal structure is illustrated in Figure 3. The structure can be approximated as a $Nd_2(NdPt)O_6$ double-perovskite where the Pt- and Nd2-sites occupy corner shared octahedra, Figure 3. However, a closer look into the crystal structure reveals that the Nd2 atom has CN = 7 and that the coordination polyhedron is actually edge-sharing to two of the $PtO_6$-octahedra. The Nd2-O polyhedron is heavily distorted, with Nd2-O distances between $d$(Nd2-O2) = 2.300(2) Å and $d$(Nd2-O6) = 2.809(2) Å. For these reasons, the structure has complexity beyond that of the regular double perovskite type structure. This is discussed in detail by Ferreira et al. in the case of $Ln_9Sr_2Ir_4O_{24}$ [13].

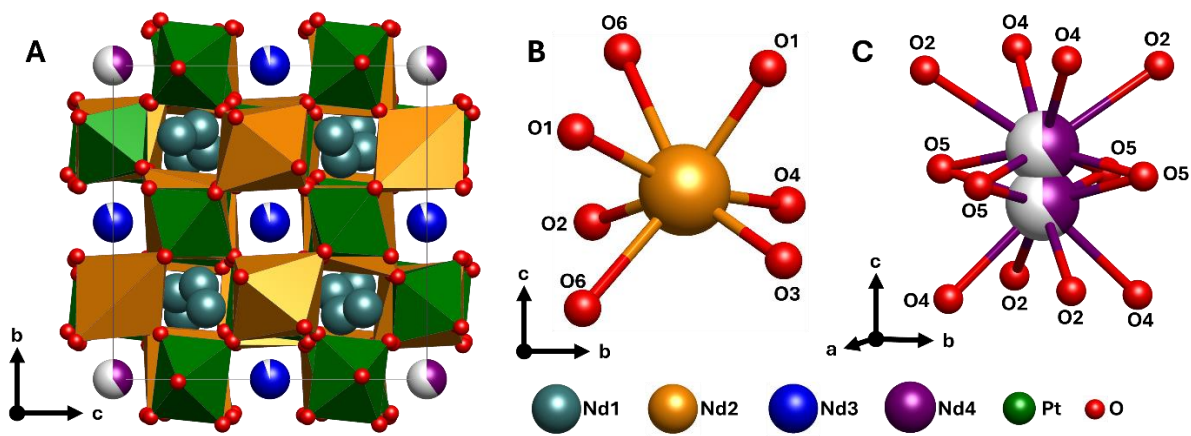

**Figure 3**. **A** – Crystal structure of $Nd_{10.67}Pt_4O_{24}$ with the neodymium sites illustrated in different colors. The oxygen coordination around the, **B** – Nd2 site, and **C** – Nd4 site. Partial occupancy as obtained from neutron powder diffraction data is illustrated with white area on the atoms Nd3 and Nd4.

The Pt-atoms take a quite regular octahedral environment, with average Pt-O distance of 2.0203 and 2.0324 Å, and distortion index ($D = \frac{1}{n}\sum_{i=1}^{n}\frac{|l_i - l_{av}|}{l_{av}}$) of 1.135 and 0.454 % for Pt1 and Pt2 respectively. These bond lengths are fully consistent with expectations for Pt(IV). Based on Shannon radii [30] one notes that Nd-Pt-O would fulfil the pyrochlore stability criterion, however, barely the Goldschmidt $t$-factor criterion for the perovskite structure (here calculated $t$ = 0.71). This structure type has been discussed in detail by Ferreira *et al.* [13], both in terms of a framework structure based on Nd1- and Nd2-polyhedra, and in terms of unique chains of edge sharing coordination polyhedra.

Solid solubility and site disorder have been reported for the iridate analogues. In $La_9(Sr_{0.925}La_{0.075})_2Ir_4O_{24}$ it was found that La and Sr are distributed over two of the $Ln$-sites, corresponding to the Nd3 and Nd4 sites in our case [13]. For the current sample, there exists no solid solubility that could provide a similar site disorder. However, for both $Nd_{10.67}Pt_4O_{24}$ and $La_9(Sr_{0.925}La_{0.075})_2Ir_4O_{24}$ the Nd4 (and Nd3)-site is just partly occupied; occupation numbers being respectively 0.33 and 0.50 [13]. This partial occupancy is required to ensure charge neutrality. For the compounds $A_{11}Re_4O_{24}$ ($A$ = Ca, Sr) [14,15] and $Ba_{11}Os_4O_{24}$ [16] with mixed valence states; Re(VI) and Re(VII), and likewise Os(VI) and Os(VII), there are no such $A$-site vacancies, Table 4. For the latter $Ba_{11}Os_4O_{24}$ [16] the relevant Ba-cation is located to the *4a*-site at (0,1/4,1/8) which corresponds to the center between the split positions in Figure 3 where Nd4 takes a partly filled *8e*-site.

**Table 5:** Formal oxidation state for the *5d*-cations in isostructural compounds, in which charge neutrality is obtained by mixed valence, whereas in $Nd_{10.67}Pt_4O_{24}$ by vacancies on Nd.

| Compound | Specie I | Specie II | Ratio | Reference |
|---|---|---|---|---|
| $A_{11}Re_4O_{24}$ (*A* = Ca, Sr) | Re(VI) | Re(VII) | 1 : 1 | [14,15] |
| $A_{11}Os_4O_{24}$ (*A* = Sr, Ba) | Os(VI) | Os(VII) | 1 : 1 | [16] |
| $La_9Sr_2Ir_4O_{24}$ | Ir(IV) | Ir(V) | 3 : 1 | [13] |
| $Nd_{10.67}Pt_4O_{24}$ | Pt(IV) | | | This work |

**Thermal stability and evaluation of the oxidation state of platinum in $Nd_{10.67}Pt_4O_{24}$**

To verify the oxidation state of platinum in $Nd_{10.67}Pt_4O_{24}$, we evaluated the thermal stability and its decomposition products by TGA sending a 33 vol. % $O_2$ in $N_2$ gas mixture over the sample. Upon heating at a rate of 20 °C/min between 25 and 750 °C, followed by a 1 °C/min ramping rate to 1100 °C, $Nd_{10.67}Pt_4O_{24}$ undergoes three mass loss events, Figure 4. A small mass loss is observed below 400 °C, which probably reflects evaporation of adsorbed water or chemisorbed water (hydroxides) and/or $CO_2$ (carbonate oxides). The second and third events have a total mass loss of 4.3 %, corresponding well to complete reduction of $Nd_{10.67}Pt_4O_{24}$ into $Nd_2O_3$, Pt metal and $O_2$, which has a theoretical mass loss of 4.4 %. Similarly, a mass loss of 4.3 % is observed for the thermal decomposition of $Nd_{10.67}Pt_4O_{24}$ in $N_2$ gas (not shown). The powder X-ray diffraction pattern of the TGA residue after decomposition at 1100 °C consists entirely of $Nd_2O_3$ and Pt; see Supporting Information Figure S8.

The TGA data for a slow heating rate of 1 °C/min above 750 °C (Figure 4) show that the onset temperature of the decomposition is at ~930 °C. Additionally, it became possible to study a distinct plateau at around 960 °C, indicating the likely presence of an intermediate phase. To isolate this intermediate phase, an experiment was carried out within the TGA apparatus where the heating was terminated after reaching the plateau weight. The powder X-ray diffraction pattern for the TGA residue, after cooling, is shown in the Supplementary Information Figure S9. Sharp characteristic diffraction peaks of Pt(s) are observed along with very broad, yet distinct diffraction features of an ill-defined product. We suggest that the product represents a two-phase mixture between Pt and an unknown Nd-enriched platinate, probably a Pt(II) phase. Indeed, Rietveld refinements assuming the presence of an $Nd_4PtO_7$ phase (not earlier reported) that takes an $La_4PtO_7$ type structure [10], are consistent with the observed powder X-ray diffraction data, see Figure S9.

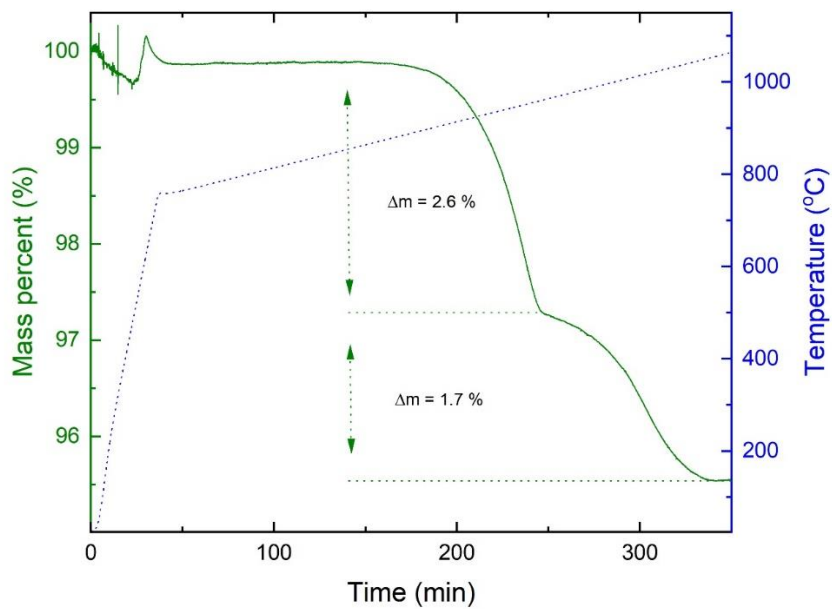

**Figure 4.** TGA of $Nd_{10.67}Pt_4O_{24}$ upon heating at a rate of 20 °C/min between 25 and 750 °C, followed by a heating rate of 1 °C/min to 1100 °C. Gas atmosphere over the sample is 33 vol. % $O_2$ in $N_2$.

**Discussion**

$Nd_{10.67}Pt_4O_{24}$ takes a similar crystal structure as the earlier reported compounds $Ln_9Sr_2Ir_4O_{24}$ [13], including $A_{11}Re_4O_{24}$ ($A$ = Ca, Sr) [14,15] and $Ba_{11}Os_4O_{24}$ [16]. In these, the *5d*-cations (Re, Ir, Os, Pt) take octahedral coordination and form along with some of the electropositive cations (*Ln* or alkaline earth) a complex double perovskite-like atomic arrangement. In several of these compounds, the 5d-cations take mixed oxidations states, with charge ordering at two octahedral sites; their formal oxidation states are listed in the Supplemental Information.

For $Nd_{10.67}Pt_4O_{24}$, platinum takes a fixed oxidation state +IV. Owing to charge neutrality we observe cation vacancies, primarily at the Nd4-sites. This imply that $Nd_{10.67}Pt_4O_{24}$ can be described as $Nd_{11-\delta}Pt_4O_{24}$ with δ = 0.33. The TGA experiments show no indications of O-non-stoichiometry below 900 °C. There is no indication of any presence of Pt(II) or a mixed valence state that could provide charge balance rather than the Nd-vacancies. Note that no vacancies occur in the isostructural rhenates and osmates (Table 5) due to a redox flexibility of the cations. On the other hand, in the iridate $Ln_9Sr_2Ir_4O_{24}$ although there is mixed states of Ir(IV) and Ir(V) cations, partially filled 8*e*-sites with *Ln* and Sr are reported, however, this just represents split positions as currently also observed. One may speculate whether a potential compound of $Nd_{10}SrPt_4O_{24}$ may exist, without cation vacancies and split positions.

Interestingly, we have noted that $Nd_{10.67}Pt_4O_{24}$ is formed when $PtO_2(g)$ reacts with $Nd_2O_3$ as described in equation 2. This reaction path does occur at elevated temperature in the Ostwald process for nitric acid/fertilizer production where gaseous $PtO_2(g)$ is lost from the Pt-Rh catalyst and in turn is captured by means of an appropriate $Nd_2O_3$ based catchment system as described in Hessevik *et al*. [11].

## Conclusion

In summary, we have prepared a new Nd platinate, $Nd_{10.67}Pt_4O_{24}$, and described its crystal structure, structural chemistry, and vacancies in detail by a combination of diffraction probes and thermogravimetry. The compound is structurally related to several complex rhenates, iridates and osmates with a double perovskite like atomic arrangement. In contrast to these compounds, charge neutrality in $Nd_{10.67}Pt_4O_{24}$ is obtained by neodymium vacancies to maintain tetravalent platinum. Pt(IV) adopts an octahedral coordination in the structure. The $Nd_{10.67}Pt_4O_{24}$ is thermally stable up to some 900 °C in 33 vol. % $O_2$ in $N_2$ gas atmosphere and is readily obtained according to a soft chemistry synthesis approach or by a direct reaction using $PtO_2(g)$. The latter is relevant during the Ostwald process for nitric acid/fertilizer production when a $Nd_2O_3$ based catchment system is used.

ASSOCIATED CONTENT

**Supporting Information** contains additional information about 3D ED refinements and Rocking curve profiles, atomic coordinates and Rietveld refinement from X-ray diffraction data, Rietveld refinement and bond lengths from neutron diffraction data, Rietveld refinement of the decomposition product and intermediate phase from TGA.

AUTHOR INFORMATION

**Corresponding Author**

Helmer Fjellvåg - *Department of Chemistry and Center for Material Science and Nanotechnology, University of Oslo, Oslo NO-0315, Norway*; Email: helmer.fjellvag@kjemi.uio.no

**Author Contributions**

The manuscript was written through contributions of all authors. All authors have given approval to the final version of the manuscript.

**Funding Sources**

The project was financed by the Research Council of Norway (NFR) project iCSI (number 237922). ØSF acknowledge funding through FLI-AVAI (number 325345). GS acknowledges CzechNanoLab Research Infrastructure supported by MEYS CR (LM2023051) and the project TERAFIT - CZ.02.01.01/00/22_008/0004594.

**Notes**

The authors declare no competing financial interest.

### Acknowledgements

The authors gratefully acknowledge Veronica A.-L. K. Killi for assistance with TGA and X-ray diffraction measurements, and the Norwegian Center for X-ray Diffraction, Scattering and Imaging (RECX) and the Swiss-Norwegian Beamlines (SNBL) at ESRF, NFR project 208896. We acknowledge ISIS neutron and muon source for Xpress beamtime at GEM and the skillful assistance of Ivan da Silva for the measurement. This work is financed through "iCSI – industrial Catalysis, Science and Innovation" center, funded by the Research Council of Norway (NFR) project no. 237922. Ø.S.F. acknowledge funding from NFR the through project 325345.

# Supplementary information: Crystal structure of $Nd_{10.67}Pt_4O_{24}$, a new neodymium platinate

*Øystein Slagtern Fjellvåg[†], Helmer Fjellvåg[‡,*], Julie Hessevik[‡], Anja Olafsen Sjåstad[‡] and Gwladys Steciuk[§]*

[†] Department for Hydrogen Technology, Institute for Energy Technology, P.O. Box 40, Kjeller NO-2027, Norway;

[‡] Center for Materials Science and Nanotechnology, Department of Chemistry, University of Oslo, N-0315 Oslo, Norway,

[§] Institute of Physics of the CAS, Na Slovance 1999/2, 182221 Prague 8, Czech Republic

**TABLES**

- Atomic coordinates from synchrotron X-ray diffraction are given in Table S1.
- Table S2 shows atomic distances from powder neutron diffraction refinements.

**FIGURES**

- The Rocking curve profiles from the 3D ED refinements are given in Figure S1.
- Figure S2 shows the Rietveld refinement of the synchrotron X-ray diffraction data.
- Figure S3-S7 show the refinements of the different detector banks of the powder neutron diffraction data.
- Figure S8 shows the Rietveld refinement of the decomposition product from TGA, which is a mix between Pt and $Nd_2O_3$.
- Figure S9 shows the Rietveld refinement of the intermediate phase refined with Pt and $La_4PtO_7$ structures.

**Table S1.** *Atomic coordinates and isotropic displacement parameters for $Nd_{10.67}Pt_4O_{24}$ based on refinement of synchrotron powder diffraction data. Space group $I4_1/a$ with lattice parameters a = 11.35203(11) Å and c = 16.21114(18) Å. Calculated standard deviations in parentheses. The refined occupation number for the Nd4 site is 0.398(5).*

| Atom | x | y | z | Uiso (Å$^2$) | Site |
|---|---|---|---|---|---|
| Nd1 | 0.72578(18) | 0.70409(15) | 0.50945(14) | 0.0113(5) | *16f* |
| Nd2 | 0.79315(14) | 0.4759(2) | 0.66280(9) | 0.0160(5) | *16f* |
| Nd3 | 0 | 0.5 | 0.49110(15) | 0.0194(6) | *8e* |
| Nd4 | 0.5 | 0.5 | 0.5341(3) | 0.010(2) | *8e* |
| Pt1 | 0.75 | 0.5 | 0.375 | 0.0067(5) | *8c* |
| Pt2 | 0 | 0.75 | 0.625 | 0.0059(4) | *8d* |
| O1 | 0.9105(15) | 0.3304(19) | 0.7148(9) | 0.0052(17) | *16f* |
| O2 | 0.6473(15) | 0.6366(18) | 0.3723(12) | 0.0052(17) | *16f* |
| O3 | 0.8877(15) | 0.6033(17) | 0.5903(10) | 0.0052(17) | *16f* |
| O4 | 0.8719(17) | 0.632(2) | 0.4015(11) | 0.0052(17) | *16f* |
| O5 | 0.502(2) | 0.7043(13) | 0.5028(7) | 0.0052(17) | *16f* |
| O6 | 0.6761(18) | 0.8880(17) | 0.4556(12) | 0.0052(17) | *16f* |

**Table S2:** Metal-oxygen bond distances from Rietveld refinements of neutron powder diffraction data at room temperature.

| Bond | Distance (Å) | Distance (Å) |
|---|---|---|
| Nd1-O1 | 2.500(2) | |
| Nd1-O2 | 2.507(2) | 2.605(2) |
| Nd1-O3 | 2.540(2) | |
| Nd1-O4 | 2.677(2) | 2.734(2) |
| Nd1-O5 | 2.430(3) | 2.471(3) |
| Nd1-O6 | 2.403(2) | |
| Nd2-O1 | 2.327(2) | 2.358(2) |
| Nd2-O2 | 2.300(2) | |
| Nd2-O3 | 2.215(2) | |
| Nd2-O4 | 2.565(2) | |
| Nd2-O5 | 2.877(2) | |
| Nd2-O6 | 2.698(2) | 2.809(2) |
| Nd3-O2 | 2.954(2) × 2 | |
| Nd3-O3 | 2.459(2) × 2 | |
| Nd3-O4 | 2.437(2) × 2 | |
| Nd3-O6 | 2.433(2) × 2 | |
| Nd4-O1 | | |
| Nd4-O2 | 2.846(3) × 2 | 3.584(4) × 2 |
| Nd4-O3 | | |
| Nd4-O4 | 2.662(3) × 2 | 3.525(4) × 2 |
| Nd4-O5 | 2.4053(19) × 4 | 2.434(2) × 4 |
| Nd4-O6 | | |
| Pt1-O2 | 1.9859(19) × 2 | |
| Pt1-O4 | 2.049(2) × 2 | |
| Pt1-O5 | 2.0254(16) × 2 | |
| Pt1-O1 | 2.0446(17) × 2 | |
| Pt1-O3 | 2.0340(19) × 2 | |
| Pt1-O6 | 2.019(2) × 2 | |

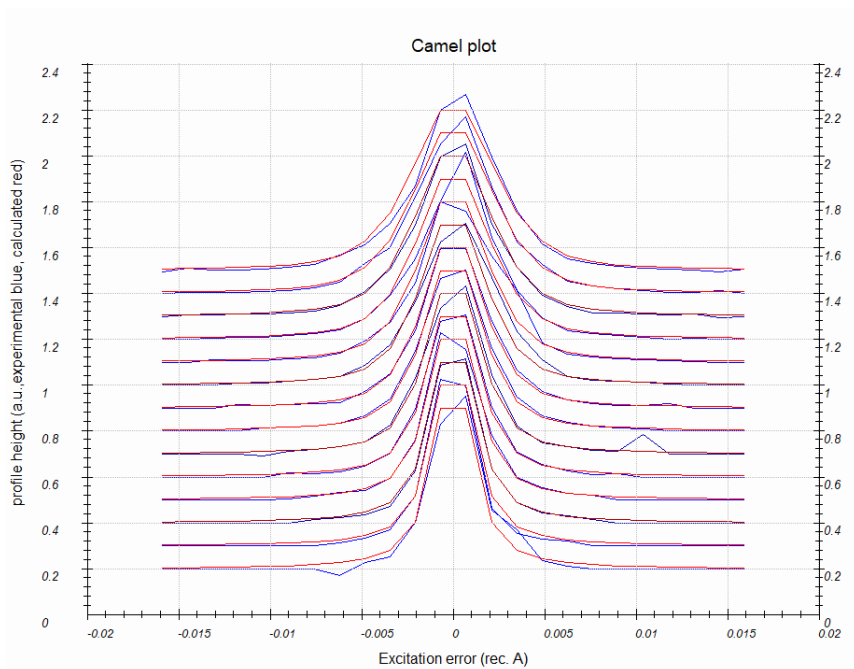

**Figure S1:** Plot of the rocking-curve profiles (Camel plot) of the experimental 3D ED data. The lowest blue curve is the average observed rocking curve in the range of 0.2 to 0.3 Å−1, and the next ones are obtained by steps of 0.1 Å−1. The red dotted curves are calculated from the *Rocking curve width* = 0.00105 Å$^{-1}$, the apparent mosaicity = 0.04796 °, and the tilt semi-angle = 0.2 °. Reflections are involved in the Camel plot for I > 5*σ(I) [31].

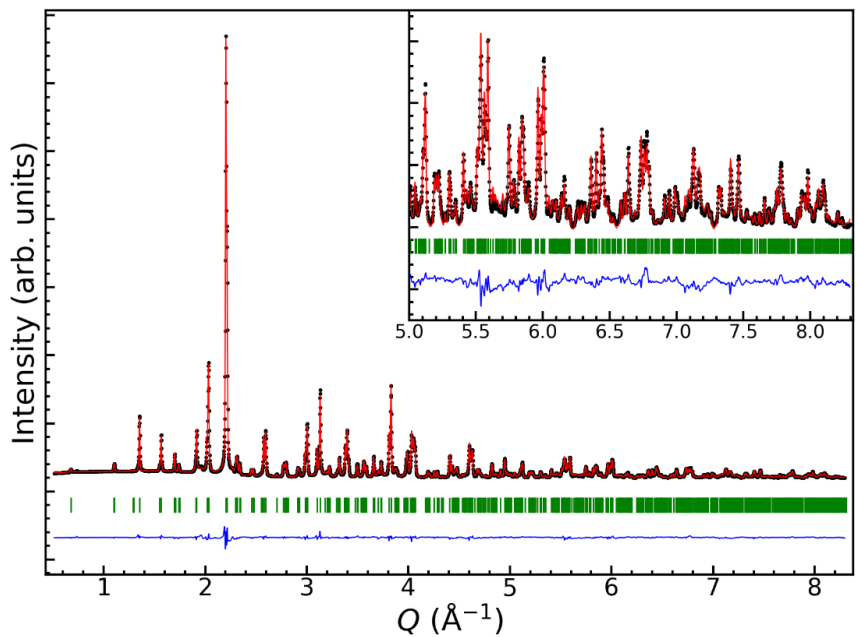

**Figure S2.** Rietveld refinement of $Nd_{10.67}Pt_4O_{24}$ from synchrotron powder diffraction data with a wavelength 0.25509 Å. The diagram shows the measured data (black dots), the calculated plot (red line), and the difference curve (blue). The reflection positions are given by green ticks. Insert show the fit to high-angle data.

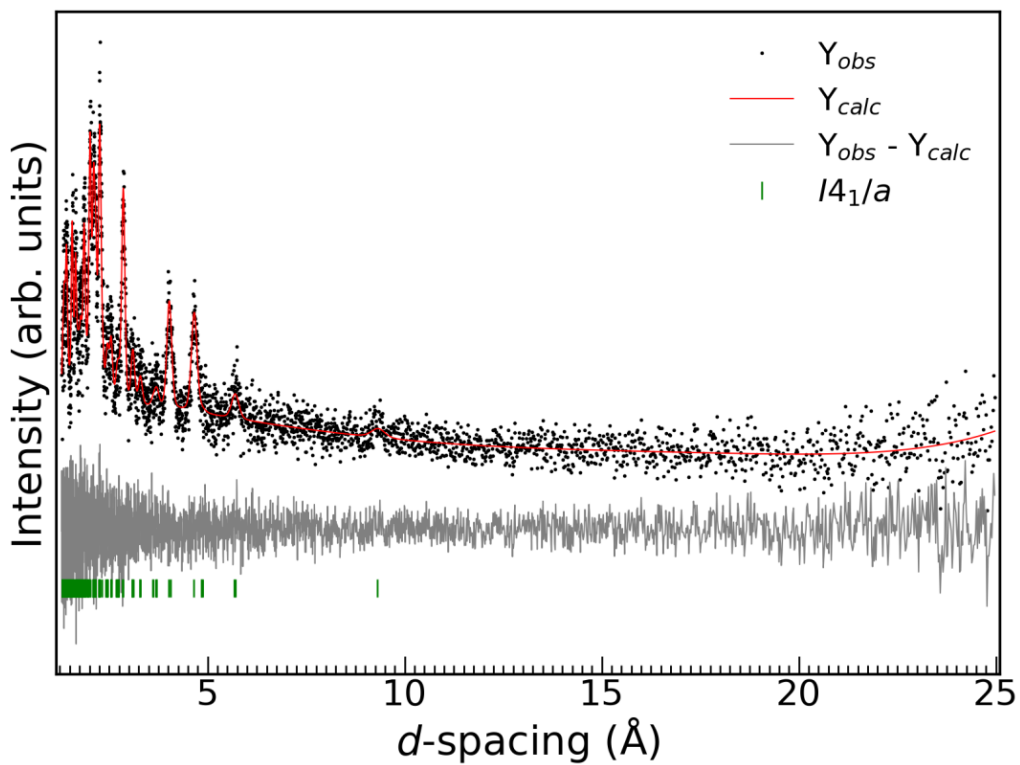

**Figure S3:** Rietveld refinement of neutron powder diffraction data from the first detector bank at GEM.

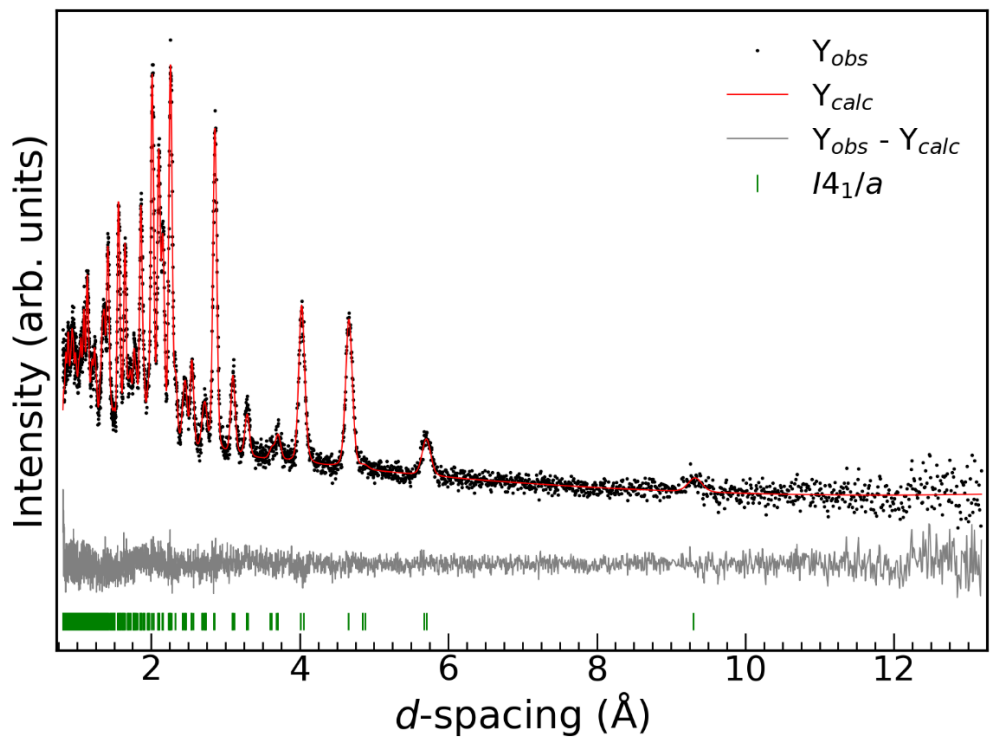

**Figure S4:** Rietveld refinement of neutron powder diffraction data from the second detector bank at GEM.

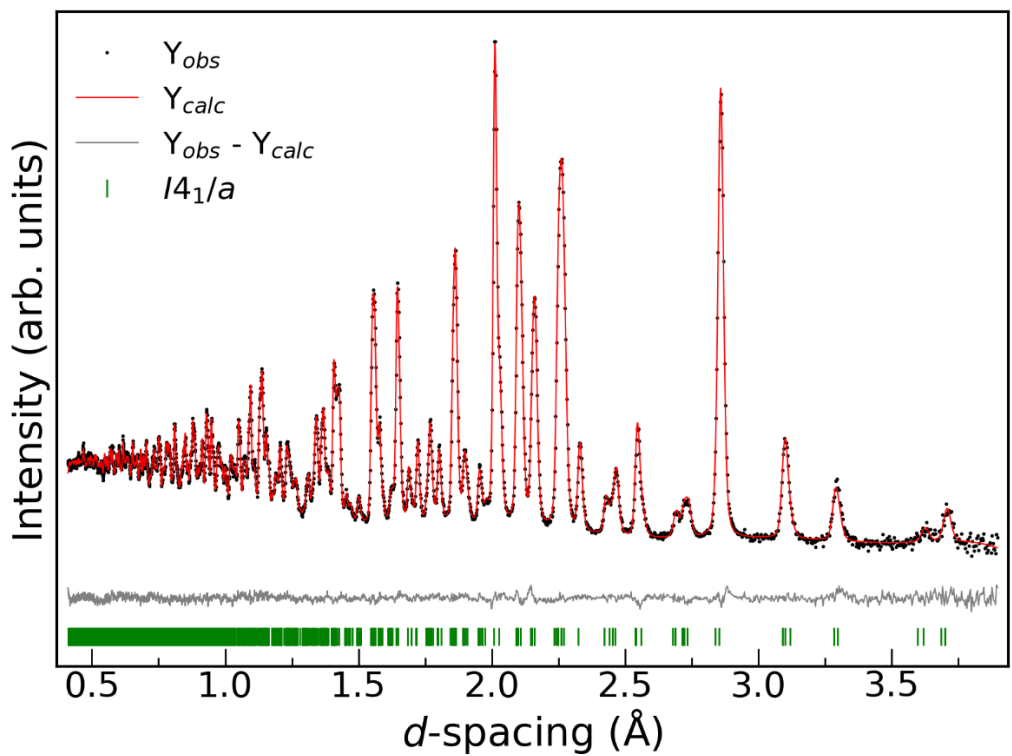

**Figure S5:** Rietveld refinement of neutron powder diffraction data from the fourth detector bank at GEM.

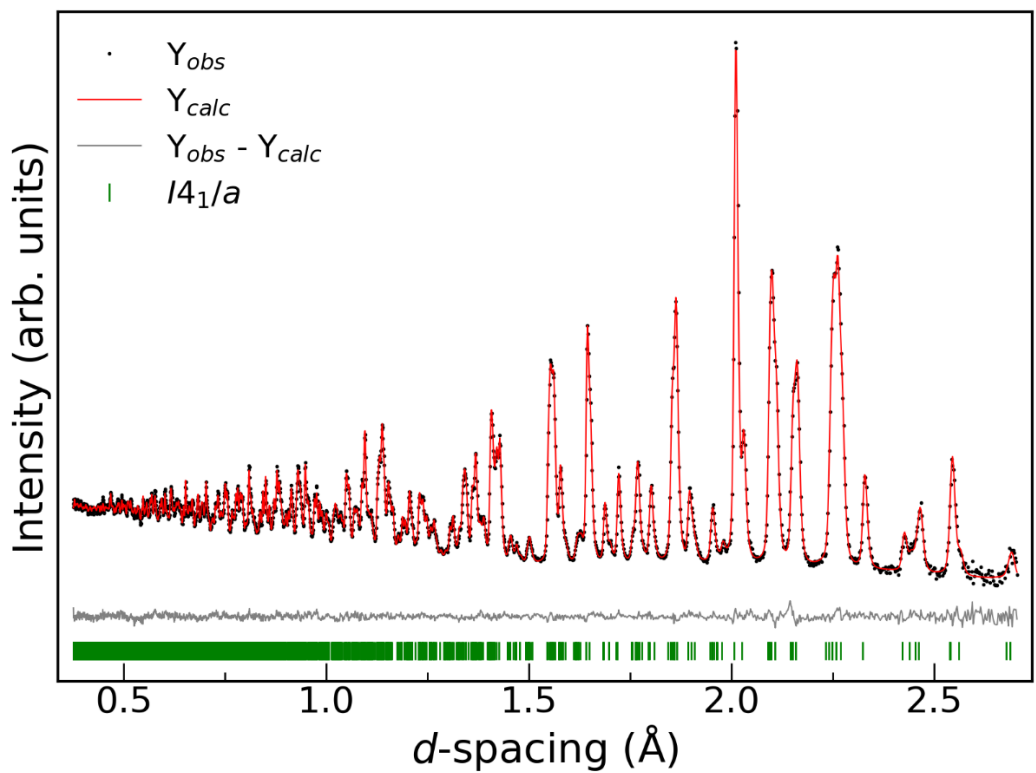

**Figure S6:** Rietveld refinement of neutron powder diffraction data from the fifth detector bank at GEM.

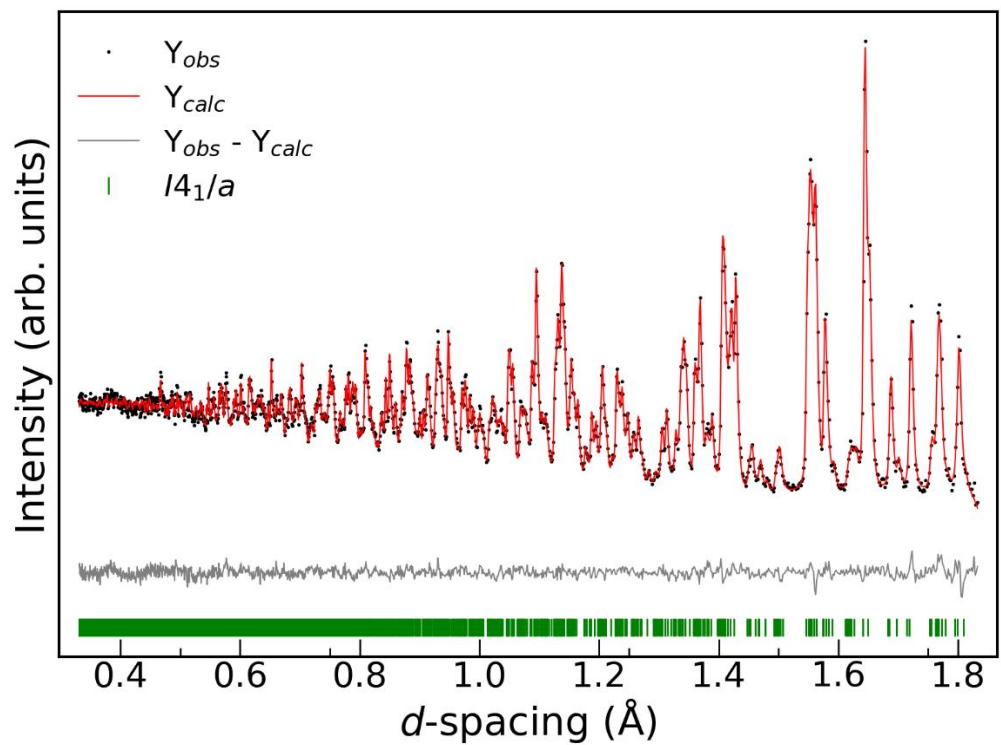

**Figure S7:** Rietveld refinement of neutron powder diffraction data from the sixth detector bank at GEM.

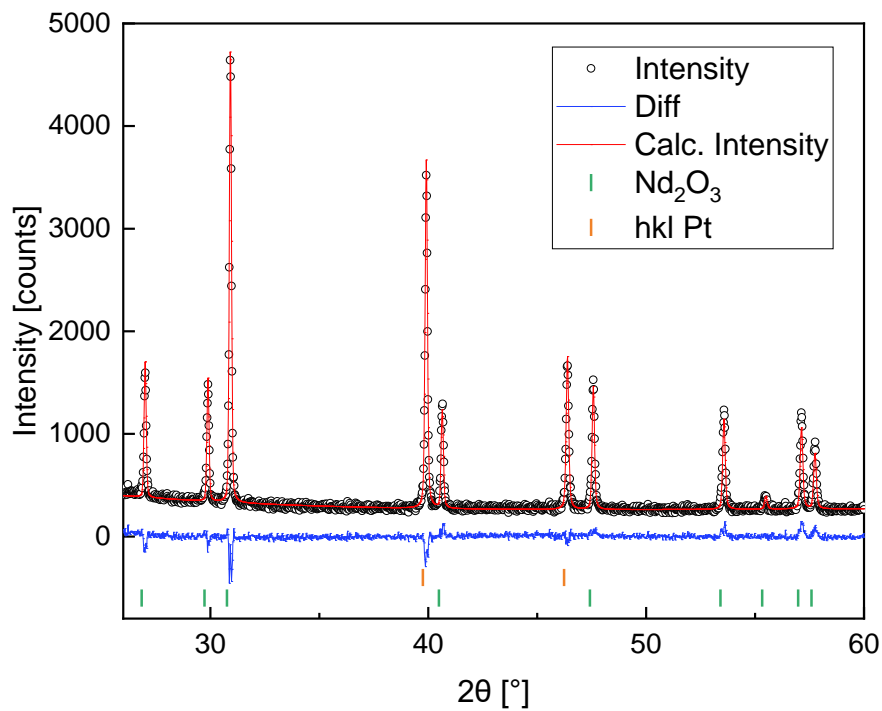

**Figure S8.** Rietveld refinement of powder X-ray diffraction data after thermogravimetric analysis of $Nd_{10.67}Pt_4O_{24}$ in a $N_2$ gas atmosphere to 1100 °C. The refinement shows a mixture of $Nd_2O_3$ and Pt.

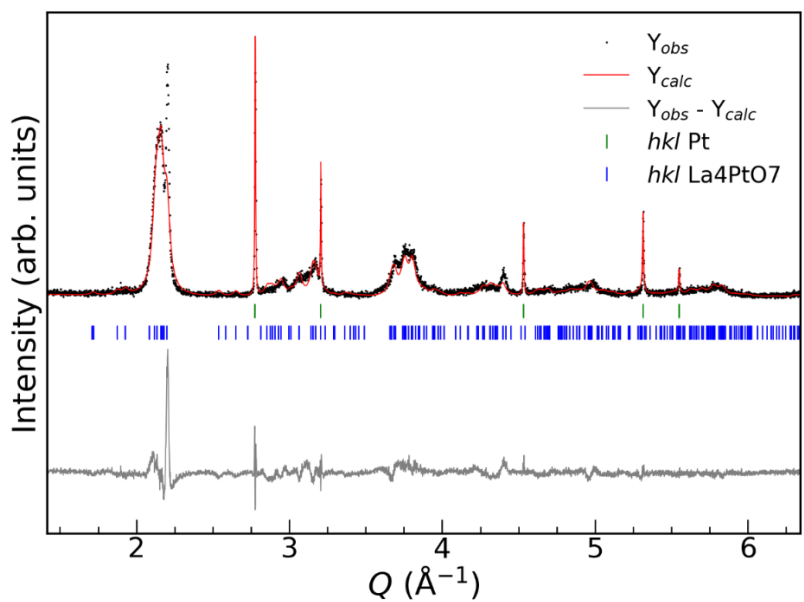

**Figure S9.** Rietveld refinement using Pt and the La$_4$PtO$_7$ phase obtained by decomposition of Nd$_{10.67}$Pt$_4$O$_{24}$ in a 33 vol. % O$_2$ in N$_2$ gas atmosphere up to 960 °C. No atomic coordinates were refined, values from Ref [1] was used. Background, sample displacement, lattice parameters and peak broadening were refined. Refined lattice parameters: Pt: $a$ = 3.9227(2) Å, La$_4$PtO$_7$: $a$ = 9.493(2) Å, $b$ = 3.9857(7) Å, $c$ = 9.223(2) Å, $β$ = 91.55(2) Å.